\documentclass[12pt]{article}

\usepackage{amssymb}
\usepackage{amsmath}

\usepackage{graphicx}

\begin{document} %


\title{A search for axion-like particles in light-by-light scattering at the CLIC}

\author{
S.C. \.{I}nan\thanks{Electronic address: sceminan@cumhuriyet.tr}
\\
{\small Department of Physics, Sivas Cumhuriyet University, 58140,
Sivas, Turkey}
\\
{\small and}
\\
A.V. Kisselev\thanks{Electronic address:
alexandre.kisselev@ihep.ru} \\
{\small A.A. Logunov Institute for High Energy Physics, NRC
``Kurchatov Institute''},
\\
{142281, Protvino, Russian Federation}}

\date{}

\maketitle

\begin{abstract}
The virtual production of axion-like particles (ALPs) in the
light-by-light scattering at the CLIC collider is studied. Both
differential and total cross sections are calculated, assuming
interaction of the ALP with photons via CP-odd term in the
Lagrangian. The 95\% C.L. exclusion regions for the ALP mass and its
coupling constant are given. By comparing our results with existing
collider bounds, we see that the ALP search at the CLIC has a great
physics potential of searching for the ALPs, especially, in the mass
region 1 TeV -- 2.4 TeV, with the collision energy $\sqrt{s} = 3000$
GeV and integrated luminosity $L = 5000$ fb$^{-1}$ for the Compton
backscattered initial photons. In particular, our limits are
stronger than recently obtained bounds for the ALP production in the
light-by-light scattering at the LHC.
\end{abstract}

\maketitle


\section{Introduction} %

The notion of the QCD axion is closely related to the strong CP
problem, which means the absence of the CP violation in the strong
interactions. In its turn, the CP problem arises as a possible
solution to the $U(1)$ problem. The QCD Lagrangian in the limit of
vanishing masses of $u$ and $d$ quarks has a global symmetry $U(2)_V
\times U(2)_A = SU(2)_I \times U(1)_Y \times SU(2)_A \times U(1)_A
$. The non-zero quark condensates $\langle \bar{u}u \rangle$ and
$\langle \bar{d}d \rangle$ break down the axial symmetry $SU(2)_A
\times U(1)_A$ spontaneously. As a result, four Nambu-Goldstone
bosons should appear. But besides light pions, no another light
state is present in the hadronic spectrum since $m_\eta \gg m_\pi$.
It is called the $U(\emph{1})$ \emph{problem} \cite{Weinberg:1975}.

The $U(1)_A$ symmetry is connected with a transformation of the
fermion fields $\psi \rightarrow e^{i\alpha \gamma_5}\psi$,
$\bar{\psi} \rightarrow \bar{\psi} e^{i\alpha \gamma_5}$. One
possible resolution of the $U(\emph{1})$ problem is provided by the
Adler-Bell-Jackiw chiral anomaly for the axial current $J_\mu^5 =
\bar{\psi} \gamma_\mu\gamma_5 \psi$ \cite{Adler:1969}.

Although the axial anomaly is a total divergency, $U(1)_A$ is not a
symmetry of the strong interaction for gluon fields $A_\mu$ which
are pure gauges at spatial infinity \cite{tHooft:1976}. They are
classified by the integer $n$, $A_\mu = (-i/g) \,\partial_\mu \omega
\omega^{-1}$, where $\omega \rightarrow \omega_n = \exp(i2\pi n)$ as
$r \rightarrow \infty$. This condition is a definition of a
classical vacuum of the gauge field $|n\rangle$. The true or
$\theta$-vacuum becomes a superposition of the vacua $|n\rangle$,
$|\theta \rangle = \sum_n \exp(-in\theta) |n\rangle$
\cite{Belavin:1975}.

As a result, an effective QCD action acquires so-called
$\theta$-term. It breaks P- ant T-invariance but conserves
C-invariance, so CP-invariance is violated. Thus, it contributes to
the neutron electric dipole moment $d_n$. The current experimental
limit $d_n < 0.021 \times 10^{-23}$ e cm \cite{PDG:2018} requires
$\theta$ to be less than $10^{-9}$. The smallness of the angle
$\bar{\theta}$ is known as \emph{strong CP problem}.

The elegant solution of the CP mystery of the SM is provided by the
Peccei-Quinn (PQ) mechanism with a new, spontaneously broken
approximate global $U(1)_{\mathrm{PQ}}$ symmetry \cite{Peccei:1977}.
As it is shown in \cite{Weiberg:1978, Wilzcek:1978} it leads to a
light neutral pseudoscalar particle, the \emph{axion} $a$, which is
the Nambu-Goldstone boson of the broken $U(1)_{\mathrm{PQ}}$
symmetry. The idea is to replace the CP-violating term $\theta$ by
the CP-conserving axion. Namely, the axion field can be redefined to
absorb the parameter $\theta$. In fact, the axion replaces the QCD
theta parameter by a dynamical quantity, thereby explaining of
non-observation of the strong CP violation. Thus, the PQ mechanism
is a compelling solution to the strong CP problem.

In the PQWW scheme \cite{Peccei:1977}-\cite{Wilzcek:1978} an extra
Higgs doublet is used, and the axion mass is related to the
electroweak symmetry breaking scale. There are two models in which
the PQ symmetry is decoupled from the electroweak (EW) scale and is
spontaneously broken. It results in axions with extremely weak
couplings (``invisible'' axion). One of the models is the KSVZ model
\cite{Kim:1979}-\cite{Shifman:1980} with one Higgs doublet in which
the axion is introduced as the phase of an EW singlet scalar field.
This scalar is coupled to an additional heavy quark, and its
coupling is induced by the interaction of the heavy quarks with
other fields. In the DFSZ model
\cite{Dine:1981}-\cite{Zhitnitsky:1980} two Higgs doublets are used,
as well as an additional EW singlet scalar. The latter is coupled to
the SM fields through its interaction with the Higgs doublets.

The axion also appears in the context of the string
theory~\cite{Svrcek:2006}-\cite{Cicoli:2012}. In the string theory,
spin-zero particles must couple to a photon field since all
couplings are defined by the expectation value of scalar fields.
This implies the existence of the P-odd term in the Lagrangian
proportional to
\begin{equation}\label{odd_term_string}
-\frac{1}{4}\,g_{a\gamma\gamma} a F_{\mu\nu}\tilde{F}^{\mu\nu}  =
g_{a\gamma\gamma} a \vec{E}\!\cdot\!\vec{B} \;,
\end{equation}
where $F_{\mu\nu}$ is the electromagnetic tensor,
$\tilde{F}_{\mu\nu} = (1/2) \varepsilon_{\mu\nu\rho\sigma}
F^{\rho\sigma}$ its dual, and $a$ is the QCD axion or axion-like
particle (ALP) \cite{Halverson:2019}. ALPs can also appear  in
theories with spontaneously broken symmetries
\cite{Masso:1997}-\cite{Bellazzini:2017} or in GUT
\cite{Rubakov:1997}. Lately, a number of new theoretical schemes
with the axion as a basic quantity was developed
\cite{Co:2019}-\cite{Gherghetta:2020}. For a review on the axions
and ALPs, see \cite{Kim:1987}-\cite{Marsh:2017} and references
therein.

Both theory and phenomenology of the axions were also studied in
large \cite{Chang:2000}-\cite{Lakic:2008} and warped
\cite{Collins:2003}-\cite{Burnier:2011} extra dimensions (EDs). In
an ED framework, the mass of the axion becomes independent of the
scale associated with the breaking of the PQ symmetry. It means that
the axion mass can be treated independently of its couplings to the
SM fields.

The very low mass and small coupling axion and/or ALP are a leading
dark matter (DM) candidate, since their properties, allow them to be
stable and difficult-to-detect. Both axions and ALPs can be produced
in the early Universe and therefore constitute most of the cold DM
in the Universe \cite{Preskill:1983}-\cite{Dine:1983} (see also
recent papers \cite{Sikivie:2008}-\cite{Basilakos:2020}). The
relevance of the QCD axion and, more generally, of ALPs in
astrophysics and cosmology is of particular interest
\cite{Marsh:2016}-\cite{Luzio:2018}. Many axion DM experiments are
in progress \cite{ADMX:2019}-\cite{QUAX:2020} (see also
\cite{Graham:2015}).

The axion phenomenology involves phenomena such as stellar
evolution, axion mediated forces, dark matter detection, axion
decays, axion-photon conversion, so-called ``light shining trough
the wall'', etc.

There is a broad experimental program aiming to search for the QCD
axion via its coupling to the SM. On the other hand, many ALP
searches assume their strong couplings to the electromagnetic term
$F_{\mu\nu} \tilde{F}^{\mu\nu}$ as in eq.~\eqref{odd_term_string}.
In terrestrial experiments, bounds on very low mass axions and small
mass axions were obtained \cite{OSCAR}-\cite{BSM}. The coupling of
the ALPs to other gauge bosons is also studied (see, for instance,
\cite{Alvarez:2019_2}). Note that the ALPs are not directly relevant
for the QCD axion. Therefore, heavy ALPs can be detected at
colliders, in particular, in a light-by-light scattering
\cite{Bauer:2017}-\cite{Beldenegro:2019}. As it was shown in
\cite{Bauer:2019}, searches at the LHC with the use of the proton
tagging technique can constrain the ALP masses in the region 0.5
TeV--2 TeV.

Compact Linear Collider (CLIC) is the linear collider that is
planned to accelerate and collide electrons and positrons at
maximally $ 3 $ TeV center-of-mass energy \cite{bra}. In the CLIC,
it is possible to obtain accelerating gradients of $100$ MV/m. Three
energy states are considered to operate CLIC at maximum efficiency
\cite{bur}. The $\sqrt{s}=380$ GeV is the first one and it is
possible to reach the integrated luminosity $L=1000$ fb$^{-1}$. This
energy stage cover Higgs boson, top quark, and gauge sectors. It is
planned to examine such SM particles with high precision \cite{dan}.
The second one has $\sqrt{s}=1500$ GeV center-of-mass energy and
$2500$ fb$^{-1}$  integrated luminosity. At this stage, it is enabled
to investigate beyond the SM physics. Also, a detailed analysis of
the Higgs boson can be made, such as the Higgs self-coupling, the
top-Yukawa coupling, and rare Higgs decay channels. \cite{lin}. The
third stage of the CLIC has a maximum center-of-mass energy value
$\sqrt{s}=3000$ GeV and integrated luminosity value $L=5000$
fb$^{-1}$. At this stage, the most precise examinations of the SM is
possible. Moreover, it is enabled to discover beyond the SM heavy
particles of mass greater than $1500$ GeV \cite{dan}. The new
physics search potential of the CLIC is presented in
\cite{CLIC_BSM}.

At the CLIC, it is possible to study $\gamma\gamma$ and $e\gamma$
collider with real photons. These $\gamma$ beams are gotten by the
Compton backscattering of laser photons off linear electron beams.
Another options for the $\gamma\gamma$ and $e\gamma$ collisions are
photon-induced processes at the CLIC. In this type of process, the
photons are emitted from the incoming electron beams. The photons
scatter at tiny angles from the beam pipe. Hence, they have very low
virtuality; that is why these photons are called ``almost-real''.

The first evidence of the subprocess $\gamma\gamma \rightarrow
\gamma\gamma$ was observed by the Collaboration ATLAS in high-energy
ultra-peripheral PbPb collisions \cite{ATLAS_ions}. The same process
was also reported by the CMS Collaboration \cite{CMS_ions}.
Recently, the Collaboration ATLAS have published the evidence of the
light-by-light scattering with the certainty of 8,2 sigma
\cite{ATLAS_ions_2}. The analysis of the exclusive and diffractive
$\gamma\gamma$ production in PbPb collisions was done in
\cite{Coelho:2020}. We have examined a possibility to constrain the
parameters of the model with a warped ED in the photon-induced
process $pp \rightarrow p\gamma\gamma p \rightarrow p'\gamma\gamma
p'$ at the LHC \cite{Inan_Kisselev:2019}. Previously, the
photon-induced processes in EDs were studied in
\cite{Atag:2009}-\cite{Atag:2010}.

In the present paper, we propose to search for the ALP $a$ in the
exclusive light-by-light scattering at the lepton collider CLIC.

In the next section differential and total cross sections are
calculated as functions of the ALP mass $m_a$ and its coupling $f$.
It enables us to estimate the CLIC exclusion regions for both types
of the initial photons.

\section{Light-by-light virtual production of ALP} %

The pseudoscalar ALP couples to the SM photons via
\begin{equation}\label{axion_photon_lagrangian}
\mathcal{L}_a = \frac{1}{2}\,(\partial_\mu a)(\partial^\mu a) -
\frac{1}{2} m_a^2 a^2 + \frac{a}{f_a^{(-)}} \,F_{\mu\nu} F^{\mu\nu}
+ \frac{a}{f_a^{(+)}} \,F_{\mu\nu} \tilde{F}^{\mu\nu} \;,
\end{equation}
were $\big(f_a^{(-)}\big)^{-1}$ and $\big(f_a^{(+)}\big)^{-1}$ are
the ALP-photon couplings in CP-odd and CP-even terms of the
interaction Lagrangian. Note that, in contrast to the true QCD
axion, the mass and couplings of the ALP are independent parameters.
In what follows, we assume that only the last term is realized in
\eqref{axion_photon_lagrangian} with $f_a^{(+)} = f$. As for a
possible contribution from the third term in
\eqref{axion_photon_lagrangian}, it is discussed in the section
Conclusions.

 The explicit expressions for the photon
spectrum are given below. The differential cross section is the
following sum of helicity amplitudes squared \cite{Beldenegro:2018}
\begin{equation}\label{diff_cs}
\frac{d\sigma}{d\Omega} = \frac{1}{128\pi^2 s} \left( |M_{++++}|^2 +
|M_{+-+-}|^2 + |M_{+--+}|^2 + |M_{++--}|^2 \right) \;.
\end{equation}
Here and below $s$, $t$ and $u$ are the Mandelstam variables of the
diphoton system. Each of the helicity amplitudes is a sum of the ALP
and SM terms,
\begin{equation}\label{ALP+SM}
M = M_a + M_{\mathrm{ew}} \;.
\end{equation}

The explicit expressions of the pure ALP amplitudes can be found in
\cite{Beldenegro:2018}. In particular,
\begin{align}\label{M++++}
\mathfrak{Re}M_{++++}^{(a)} &= - \frac{4}{f_a^2} \frac{s^2(s -
m_a^2)}{(s - m_a^2)^2 + m_a^2 \Gamma_a^2} \;, \nonumber \\
\mathfrak{Im}M_{++++}^{(a)} &= \frac{4}{f_a^2} \frac{s^2 m_a
\Gamma_a}{(s - m_a^2)^2 + m_a^2 \Gamma_a^2} \;,
\end{align}
where $\Gamma_a$ is the total width of the ALP
\begin{equation}\label{axion_width}
\Gamma_a =
\frac{\Gamma(a\rightarrow\gamma\gamma)}{\mathrm{Br}(a\rightarrow\gamma\gamma)}
\;,
\end{equation}
and
\begin{equation}\label{photon_axion_width}
\Gamma(a\rightarrow\gamma\gamma) = \frac{m_a^3}{4\pi f^2}
\end{equation}
is its decay width into two photons. Correspondingly, we have
\cite{Beldenegro:2018}
\begin{equation}\label{M+-+-}
\mathfrak{Re}M_{+-+-}^{(a)} = - \frac{4}{f_a^2} \frac{u^2}{u -
m_a^2} \;, \quad \mathfrak{Im}M_{+-+-}^{(a)} = 0 \;,
\end{equation}
\begin{equation}\label{M+--+}
\mathfrak{Re}M_{+--+}^{(a)} = - \frac{4}{f_a^2} \frac{t^2}{t -
m_a^2} \;, \quad \mathfrak{Im}M_{+--+}^{(a)} = 0 \;,
\end{equation}
\begin{align}\label{M++--}
\mathfrak{Re} M_{++--}^{(a)} &=  \frac{4}{f_a^2} \left( \frac{s^2(s
- m_a^2)}{(s - m_a^2)^2 + m_a^2 \Gamma_a^2} + \frac{t^2}{t - m_a^2}
+ \frac{u^2}{u -
m_a^2}  \right) , \nonumber \\
\mathfrak{Im}M_{++--}^{(a)} &= -\frac{4}{f_a^2}\frac{s^2 m_a
\Gamma_a}{(s - m_a^2)^2 + m_a^2 \Gamma_a^2} \; ,
\end{align}
\begin{equation}\label{zero_matrix_element}
M_{+++-}^{(a)} = 0 \;.
\end{equation}
An account of the ALP width $\Gamma_a$ is mainly important in a
vicinity of the point $s \sim m_a^2$. That is why, it is omitted in
the denominators in eqs.~\eqref{M+-+-}, \eqref{M+--+}, as well as in
the last two terms in the first row of eq.~\eqref{M++--}.

The SM (electroweak) amplitude is a sum of the fermion and $W$ boson
one-loop amplitudes
\begin{equation}\label{f+W_ew}
M_{\mathrm{ew}} = M^f + M^W \;.
\end{equation}
The amplitudes $M_{++++}^f(s,t,u)$ and $M_{++++}^W(s,t,u)$ are
calculated in \cite{Jikia:1994}-\cite{Gounaris:1999} (see also
\cite{Atag:2010})
\begin{align}\label{W_f_++++}
\frac{1}{\alpha_{\mathrm{em}}^2 e_f^4}
\mathfrak{Re}M_{++++}^f(s,t,u) = &-8 -8 \left( \frac{u - t}{s}
\right) \ln\left(\frac{u}{t}\right)
\nonumber \\
&- 4 \left( \frac{t^2+u^2}{s^2} \right) \left[
\ln^2\left(\frac{u}{t}\right) + \pi^2 \right] ,
\nonumber \\
\mathfrak{Im}M_{++++}^f(s,t,u) = 0 \;,
\end{align}
where $e_f$ is the fermion electric charge in units of the proton
charge,
\begin{align}\label{M_W_++++}
\frac{1}{\alpha_{\mathrm{em}}^2} \mathfrak{Re}M_{++++}^W(s,t,u) &=
12 + 12 \left( \frac{u - t}{s} \right) \ln\left(\frac{u}{t}\right)
\nonumber \\
&+ 16 \left( 1 - \frac{3tu}{4s^2} \right) \left[
\ln^2\left(\frac{u}{t}\right) + \pi^2 \right]
\nonumber \\
&+ 16 \bigg[ \frac{s}{t} \ln\left(\frac{s}{m_W^2}\right)
\ln\left(\frac{-t}{m_W^2}\right)
+ \frac{s}{u} \ln\left(\frac{s}{m_W^2}\right) \ln\left(\frac{-u}{m_W^2}\right)
\nonumber \\
&+ \frac{s^2}{tu} \ln\left(\frac{-t}{m_W^2}\right)
\ln\left(\frac{-u}{m_W^2}\right)  \bigg] \;,
\nonumber \\
\frac{1}{\alpha_{\mathrm{em}}^2} \mathfrak{Im}M_{++++}^W(s,t,u) &=
-16\pi \left[ \frac{s}{t} \ln\left(\frac{-t}{m_W^2}\right) +
\frac{s}{u} \ln\left(\frac{-u}{m_W^2}\right) \right] .
\end{align}

The amplitudes $M_{+-+-}^{f,W}(s,t,u)$ and $M_{+--+}^{f,W}(s,t,u)$
can be obtained with the use of the following relations
\begin{align}\label{M_relations}
M_{+-+-}(s,t,u) &= M_{++++}(u,t,s) \;,
\nonumber \\
M_{+--+}(s,t,u) &= M_{++++}(t,s,u) = M_{++++}(t,u,s) \;.
\end{align}
Note that $M_{++++}(s,t,u) = M_{++++}(s,u,t)$, since it depends only
on $s$. In particular, we get
\begin{align}\label{M_+-+-}
\frac{1}{\alpha_{\mathrm{em}}^2 e_f^4}
\mathfrak{Re}M_{+-+-}^f(s,t,u) &= -8 -8 \left( \frac{s - t}{u}
\right) \ln\left(\frac{s}{-t}\right)
\nonumber \\
&- 4 \left[ \left( \frac{t^2 + s^2}{u^2} \right)
\ln^2\left(\frac{s}{-t}\right) + \pi^2 \right] ,
\nonumber \\
\frac{1}{\alpha_{\mathrm{em}}^2 e_f^4}
\mathfrak{Im}M_{+-+-}^f(s,t,u) &= 8\pi \left[ \frac{s-t}{u} +
\frac{t^2+s^2}{u^2} \ln\left(\frac{s}{-t}\right) \right] ,
\end{align}
and
\begin{align}\label{M_f_+--+}
\frac{1}{\alpha_{\mathrm{em}}^2 e_f^4}
\mathfrak{Re}M_{+--+}^f(s,t,u) &= -8 -8 \left( \frac{u - s}{t}
\right) \ln\left(\frac{-u}{s}\right)
\nonumber \\
&- 4 \left[ \left( \frac{s^2 + u^2}{t^2} \right)
\ln^2\left(\frac{-u}{s}\right) + \pi^2 \right] \;,
\nonumber \\
\frac{1}{\alpha_{\mathrm{em}}^2 e_f^4}
\mathfrak{Im}M_{+--+}^f(s,t,u) &= -8\pi \left[ \frac{u-s}{t} +
\frac{s^2+u^2}{t^2} \ln\left(\frac{-u}{s}\right) \right] .
\end{align}

The explicit formulas for $M_{+-+-}^W(s,t,u)$ have been already
derived in \cite{Atag:2010}
\begin{align}\label{M_W_+--+}
\frac{1}{\alpha_{\mathrm{em}}^2} \mathfrak{Re}M_{+-+-}^W(s,t,u) &=
12 + 12 \left( \frac{s - t}{u} \right) \ln\left(\frac{s}{-t}\right)
\nonumber \\
&+ 16 \left( 1 - \frac{3ts}{4u^2}\right) \ln^2\left(\frac{s}{-t}\right)
\nonumber \\
&+ 16 \bigg[ \frac{u}{t} \ln\left(\frac{-u}{m_W^2}\right)
\ln\left(\frac{-t}{m_W^2}\right) + \frac{u}{s}
\ln\left(\frac{-u}{m_W^2}\right) \ln\left(\frac{s}{m_W^2}\right)
\nonumber \\
&+ \frac{u^2}{ts} \ln\left(\frac{-t}{m_W^2}\right)
\ln\left(\frac{s}{m_W^2}\right) \bigg] \;,
\nonumber \\
\frac{1}{\alpha_{\mathrm{em}}^2} \mathfrak{Im}M_{+-+-}^W(s,t,u) &= -
\pi\bigg[ 12 \left( \frac{s - t}{u} \right) + 32 \left( 1 -
\frac{3ts}{4u^2}\right) \ln\left(\frac{s}{-t}\right)
\nonumber \\
&+ 16 \frac{u}{s} \ln\left(\frac{-u}{m_W^2}\right) + 16
\frac{u^2}{ts} \ln\left(\frac{-t}{m_W^2}\right) \bigg] .
\end{align}
The explicit expressions for $M_{++--}^f(s,t,u)$ and
$M_{++--}^W(s,t,u)$ are also known
\cite{Jikia:1994}-\cite{Gounaris:1999}
\begin{align}\label{M_++--}
\mathfrak{Re}M_{++--}^f(s,t,u) &= 8\alpha_{\mathrm{em}}^2 e_f^4,
\quad
\mathfrak{Im}M_{++--}^f(s,t,u) = 0 \;,
\nonumber \\
\mathfrak{Re}M_{++--}^W(s,t,u) &= -12 \alpha_{\mathrm{em}}^2, \quad
\mathfrak{Im}M_{++--}^W(s,t,u) = 0 \;.
\end{align}
Finally, neglecting terms $m_f^2/s$, $m_f^2/t$ and $m_f^2/u$, we
have
\begin{align}\label{M_+++-}
M_{+++-}^f(s,t,u) &\simeq M_{++--}^W(s,t,u) \;, \\
M_{+++-}^W(s,t,u) &\simeq M_{++--}^W(s,t,u) \;.
\end{align}

\subsection{Compton backscattered photons} %

In addition to $e^+e^-$ collisions, $e \gamma$ and $\gamma\gamma$
interactions with real photons can be examined at the CLIC. For this
purpose, real photons could be constructed by the Compton
backscattering of laser photons off linear electron beam. In this
process, a laser photon which has $E_{0}$ energy interacts with an
electron. This electron has very high energy $E_{e}$, and it collides
with the laser photon at a tiny collision angle $\alpha$ (almost
head-on). The scattered photon energy $E_{\gamma}$ can be determined
as follows \cite{Ginzburg:1981}
\begin{equation}\label{CB_spectrum}
E_{\gamma}=\frac{x_{\max}E_{e}}{1+(\theta/\theta_{0})^2}, \;, \quad
x_{\max}=\frac{\zeta}{\zeta+1} \;, \quad
\theta_{0}=\frac{m_{e}}{E_{e}}\sqrt{\zeta+1} \;.
\end{equation}
Here $\theta$ is the photon scattering angle from an
initial direction of the electron, $x_{\max}=E_{\gamma,\max}/E_{e}$,
where $E_{\gamma,\max}$ is the maximum energy of the scattered
photons, and
\begin{equation}\label{zeta}
\zeta=\frac{4E_{e}E_{0}}{m_e^2} \, \left[ \cos \left(
\frac{\alpha}{2} \right) \right]^2 \simeq \frac{4E_{e}E_{0}}{m_e^2}
= 15.3 \left( \frac{E_{e}}{\mathrm{TeV}} \right) \!\left(
\frac{E_{0}}{\mathrm{eV}} \right).
\end{equation}

The photon energy $E_{\gamma}$ grows, with an increase of $\zeta$.
Hence the energy spectrum gets to be narrower. At the same time, if
$\zeta$ value is greater than $4.8$; high energy photons may be lost
due to $e^+e^-$ pair creation in interactions of unscattered laser
photons with backscattered photons. The invariant energy of the
$\gamma\gamma$-system is equal to $s = 4x_1 x_2 E_e^2 =
4E_{\gamma_1}E_{\gamma_2}$, $x_i = E_{\gamma_i}/E_e$, where
$E_{\gamma_i}$ ($i=1,2$) is defined by eqs.~\eqref{CB_spectrum},
\eqref{zeta}.

\begin{figure}[htb]
\begin{center}
\includegraphics[scale=0.35]{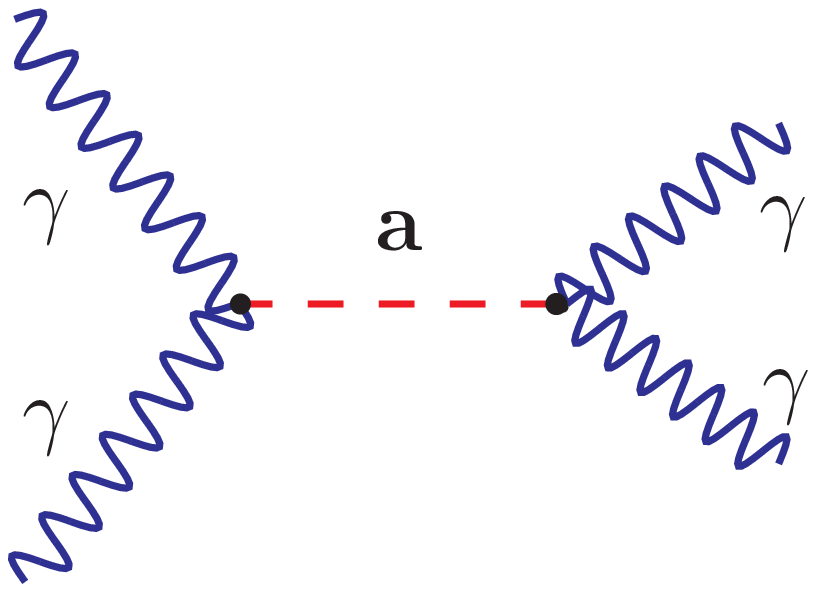}
\includegraphics[scale=0.35]{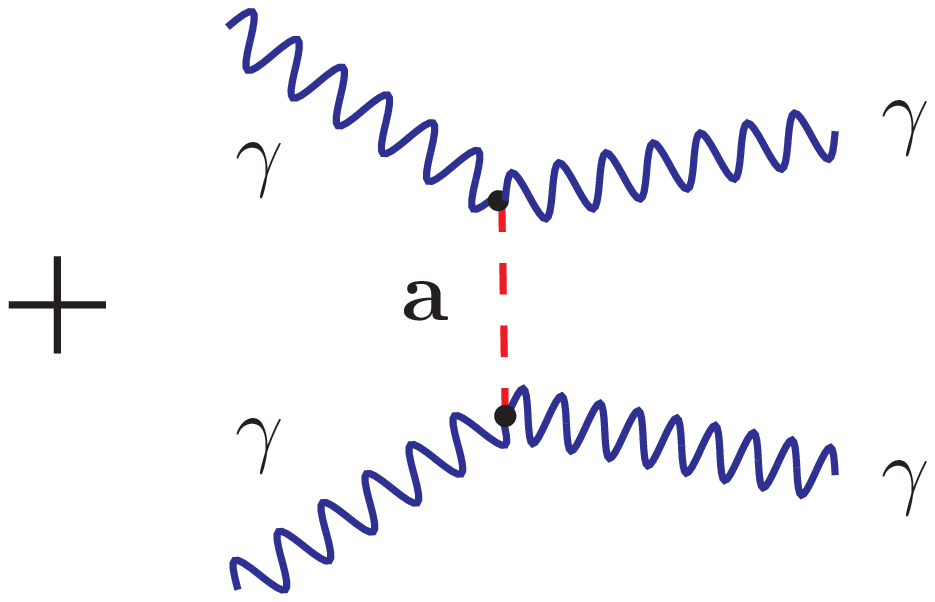}
\includegraphics[scale=0.35]{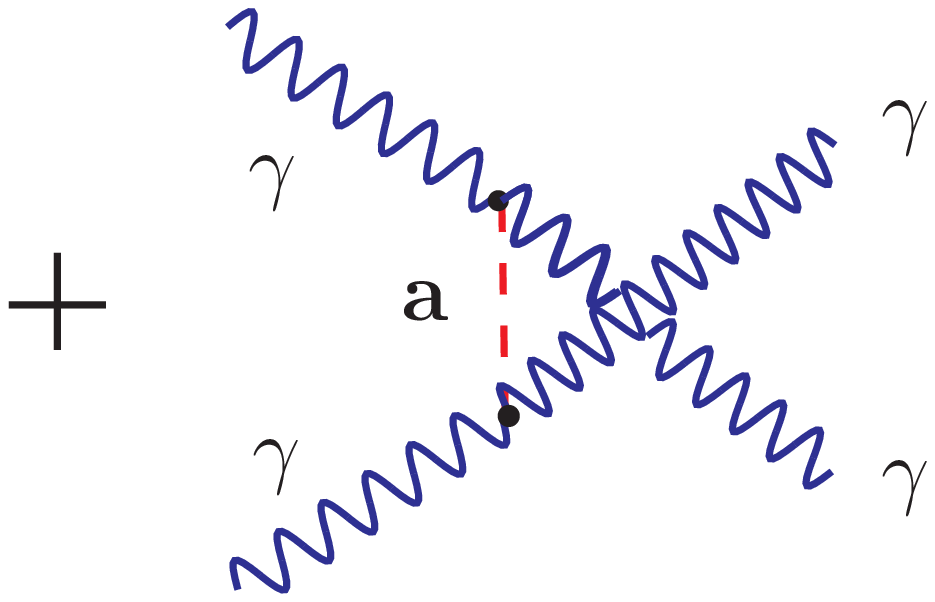}
\caption{The Feynman diagrams describing light-by-light
virtual production of the axion-like particle $a$.}
\label{fig:diag}
\end{center}
\end{figure}

Most of these real scattered photons have high energy. These Compton
backscattered (CB) photons give a spectrum which is defined as
follows \cite{Milburn:1963}-\cite{Ginzburg:1981}
\begin{equation}\label{photon_spectrum}
f_{\gamma/e}(x) = \frac{1}{g(\zeta)}\left[1-x+\frac{1}{1-x} -
\frac{4x}{\zeta(1-x)} + \frac{4x^{2}}{\zeta^{2}(1-x)^{2}}\right] ,
\end{equation}
where
\begin{equation}\label{g}
g(\zeta) =
\left(1-\frac{4}{\zeta}-\frac{8}{\zeta^2}\right)\log{(\zeta+1)} +
\frac{1}{2} + \frac{8}{\zeta}-\frac{1}{2(\zeta+1)^2} \;.
\end{equation}
Note that $x_{\max}$ reaches $0.83$ when $\zeta=4.8$. We will use
the cut on the final state photon rapidity $|\eta_{\gamma\gamma}| <
2.5$. The cross section of the diphoton production with the CB
photon at the CLIC can be found as the integration,
\begin{equation}\label{cs}
d\sigma = 2\int\limits_{z_{\min}}^{z_{\max}} \!\!dz \,z
\!\!\int\limits_{z^2/y_{\max}}^{y_{\max}} \!\!\frac{dy}{y}
\,f_{\gamma/e}(y) \,f_{\gamma/e}(z^2/y) \,d\sigma (\gamma\gamma
\rightarrow \gamma\gamma) \;.
\end{equation}
Here
\begin{equation}\label{y_z_limits}
y_{\max} = z_{\max} = 0.83 \;, \quad  z_{\min} =
\frac{p_\bot}{E_e}\;,
\end{equation}
where $p_{\bot}$ is the transverse momentum of the final photons,
$f_{\gamma/e}(y)$ is the photon spectrum, and $d\sigma (\gamma\gamma
\rightarrow \gamma\gamma)$ is the unpolarized differential cross
section of the subprocess $\gamma\gamma \rightarrow \gamma\gamma$.
The Feynman diagrams for this process are shown in
Fig.~\ref{fig:diag}. Let us note that in our calculations, we take
into account $W$-loop and fermion-loop contributions as the main SM
background.  The other possible background with fake photons from
decays of $\pi^0$, $\eta$, and $\eta'$ is negligible in the signal
region.

\begin{figure}[htb]
\begin{center}
\includegraphics[scale=0.65]{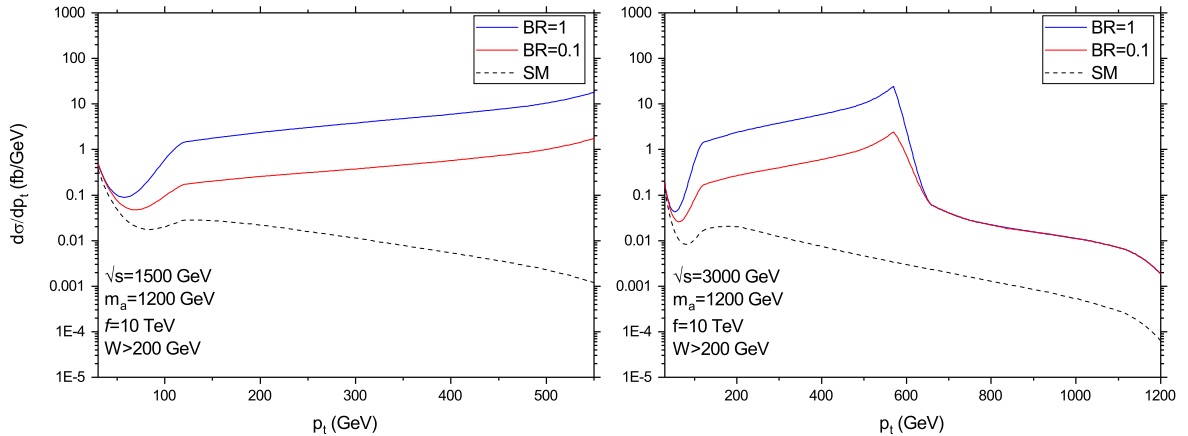}
\caption{The differential cross sections for the process
$\gamma\gamma \rightarrow \gamma\gamma$ at the CLIC for the CB
initial photons with the ALP mass $m_a = 1200$ GeV, coupling
constant $f = 10$ TeV, and cut $W>200$ GeV imposed on the photon
invariant mass $W$. The invariant energy is equal to $\sqrt{s} =
1500$ (3000) GeV in the left (right) panel. The curves both for
$\mathrm{Br}(a \rightarrow \gamma\gamma) = 1.0$ and $\mathrm{Br}(a
\rightarrow \gamma\gamma) = 0.1$ are shown. The dashed lines denote
the SM contributions.} \label{fig:CBPTD}
\end{center}
\end{figure}

\begin{figure}[htb]
\begin{center}
\includegraphics[scale=0.65]{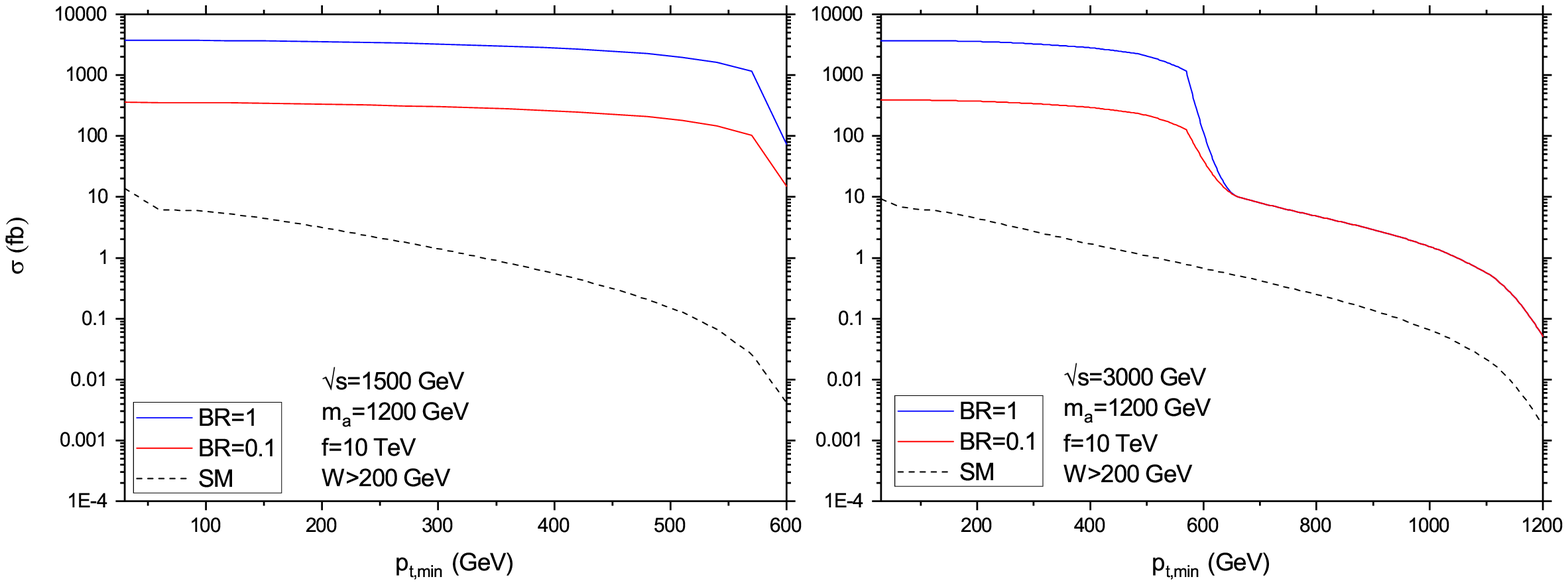}
\caption{The same as in Fig.~\ref{fig:CBPTD}, but for the total
cross sections as functions of the transverse momenta cutoff
$p_{\mathrm{t,min}}$ of the final photons.}
\label{fig:CBPTCUTWCUT}
\end{center}
\end{figure}

The differential cross sections for the process $\gamma\gamma
\rightarrow \gamma\gamma$ for the CB initial photons is shown in
Fig.~\ref{fig:CBPTD} as functions of the transverse momenta of the
final photons $p_t$. The ALP mass $m_a$ and its coupling $f$ are
chosen to be equal to 1200 GeV and 10 TeV, respectively. In order to
reduce the SM background, we have imposed the cut $W =
m_{\gamma\gamma}
> 200$ GeV. The curves are presented for two values of the ALP
branching into two photons $\mathrm{Br} = \mathrm{Br}(a \rightarrow
\gamma\gamma)$. For this differential cross sections, the virtual
production of the ALP dominates the SM light-by-light subprocess for
$p_t > 100$ GeV region. The total cross sections $\sigma(p_t >
p_{t,\min})$ as functions of the minimal transverse momenta of the
final photons $p_{t,\min}$ are shown in Fig.~\ref{fig:CBPTCUTWCUT}.
It can be seen from this figure that the deviation from the SM gets
higher as the $p_t$-cut increases. Moreover, while the SM cross
section decreases up to the value of $p_{t,\min}=500$ GeV, the total
cross section remains almost unchanged.

%

Fig.~4 demonstrates the dependence of the total cross sections on
the ALP mass for two fixed values of the ALP coupling $f=10$ TeV (in
the left panel) and $f=100$ TeV (in the right panel). Since in the
mass region $m_a = 1000-2500$ GeV, the dominant dependence of the
cross sections scales as $1/f^2$, one can assume that it comes from
the interference piece. However, it is not a case. Our calculations
have shown that the pure axion matrix element $M_a$ in (6) is
dominant. Moreover, $t$- and $u$-channel terms in Eqs.~(10)-(12)
scale as $1/f^4$ and are negligible with respect to $s$-channel
terms in Eqs.~(7) and (12). The main contribution comes from the
resonance region $s \sim m_a^2$ in the $s$-channel terms of $M_a$.
As a result, in the mass region $m_a = 1000-2500$ GeV we get
\begin{equation}\label{cross_section_appr}
\sigma \sim \frac{1}{f^2} \,\mathrm{Br(\gamma\gamma \rightarrow a)}
\;,
\end{equation}
in a qualitative agreement with Fig.~\ref{fig:CBMAE1500}. For more
details, see Appendix~A.

One can also see that the cross sections are very sensitive to the
parameter $m_a$ in the interval $m_a = 1000-2500$ GeV, in which it
is approximately two orders of magnitude greater than for $m_a$
outside of this mass range. It is not surprising that this is the
region where the value of the ALP coupling constant $f$ is mostly
restricted by the light-by-light process, see
Figs.~\ref{fig:CBSS750}, \ref{fig:CBSS1500}. In these figures, we
have applied the cut $p_t > 500$ GeV in order to suppress SM cross
sections relative to total cross sections as we analyzed from the
Fig.~\ref{fig:CBPTCUTWCUT}. In this analysis, we have used the
following statistical significance ($SS$) formula \cite{SS},
\begin{equation}\label{SS_def}
SS = \sqrt{2[(S+B) \,\ln(1 + S/B) - S]} \;.
\end{equation}
Here $S$ and $B$ are the numbers of the signal and background
events, respectively. It can be obtained that $SS \simeq S/\sqrt{B}$
for $S \ll B$. It is assumed that the uncertainty of the background
is negligible.

Our obtained exclusion regions should be compared with the current
exclusion regions on the ALP coupling and ALP mass presented in
Fig.~\ref{fig:balfig7}, especially with that obtained for the
process $pp \rightarrow p(\gamma\gamma\rightarrow\gamma\gamma)p$ at
the LHC \cite{Beldenegro:2018}. This comparison demonstrates the
great potential of the light-by-light scattering at the CLIC. Our
$95\%$ C.L. parameter exclusion region is presented in
Fig.~\ref{fig:CBSS750} for $\sqrt{s}=1500$ GeV and $L=2500$
fb$^{-1}$ using $\mathrm{Br}(a \rightarrow \gamma\gamma) = 1.0$,
0.5, and 0.1. The best bounds are achieved for $\mathrm{Br}(a
\rightarrow \gamma\gamma) = 1.0$. This figure shows the upper bound
$f^{-1} < 5.5\times 10^{-2}$ TeV$^{-1}$ for the ALP mass interval 10
GeV--800 GeV, while the light-by-light scattering at the LHC gives
the bound $f^{-1} < 4\times 10^{-1}$ TeV$^{-1}$ for the same mass
interval. Moreover, we have obtained the very strong upper bound on
$f^{-1}$, which is of the order of $10^{-4}$ TeV$^{-1}$ for the mass
range $m_a = 1000-1200$ GeV. The best limit for the $pp \rightarrow
p(\gamma\gamma\rightarrow\gamma\gamma)p$ is of the order of
$10^{-2}$ TeV$^{-1}$ for the mass range $m_a=600-800$ GeV, as seen
from Fig.~\ref{fig:balfig7}. The 95\% C.L. exclusion region for
$\sqrt{s}=3000$ GeV and $L=5000$ fb$^{-1}$ is presented in
Fig.~\ref{fig:CBSS1500}. It demonstrates the wider exclusion
regions. In particular, one can be derived the upper bound $f^{-1} <
3\times 10^{-2}$ TeV$^{-1}$ for the ALP mass interval 10 GeV--800
GeV. The stronger bounds on $f^{-1}$ have been obtained, which are of
the order of $10^{-4}$ TeV$^{-1}$ for the mass range $m_a =
1000-2400$ GeV and $\mathrm{Br}(a \rightarrow \gamma\gamma) = 1.0$.

\begin{figure}[htb]
\begin{center}
\includegraphics[scale=0.65]{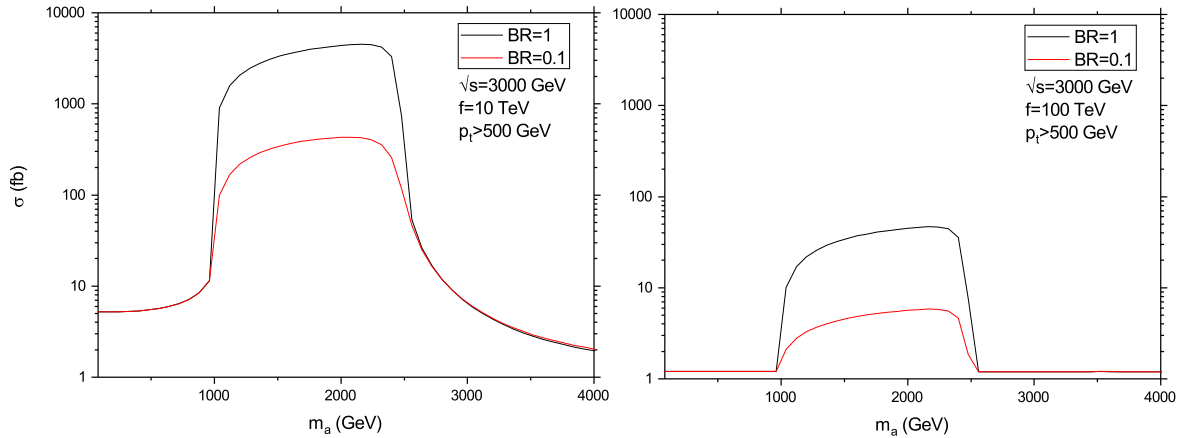}
\caption{The total cross sections for the process $\gamma\gamma
\rightarrow \gamma\gamma$ at the CLIC for the CB initial photons as
functions of the ALP mass $m_a$ for $f = 10$ TeV and $f= 100$ TeV
with two values of $\mathrm{Br}(a \rightarrow \gamma\gamma)$.}
\label{fig:CBMAE1500}
\end{center}
\end{figure}
%
\begin{figure}[htb]
\begin{center}
\includegraphics[scale=0.55]{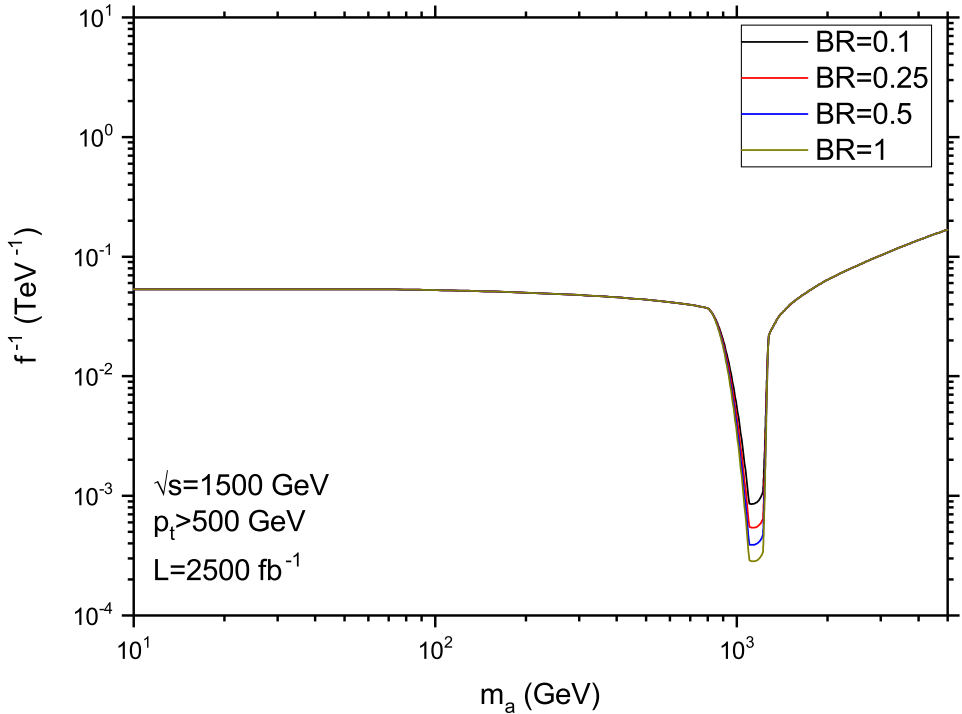}
\caption{The 95\% C.L. CLIC exclusion region for the process
$\gamma\gamma \rightarrow \gamma\gamma$ for the CB initial photons
with the invariant energy $\sqrt{s} = 1500$ GeV, cut $W>200$ GeV on
the photon invariant mass, integrated luminosity $L = 2500$
fb$^{-1}$, and different values of $\mathrm{Br}(a \rightarrow
\gamma\gamma)$. } \label{fig:CBSS750}
\end{center}
\end{figure}
%
\begin{figure}[htb]
\begin{center}
\includegraphics[scale=0.55]{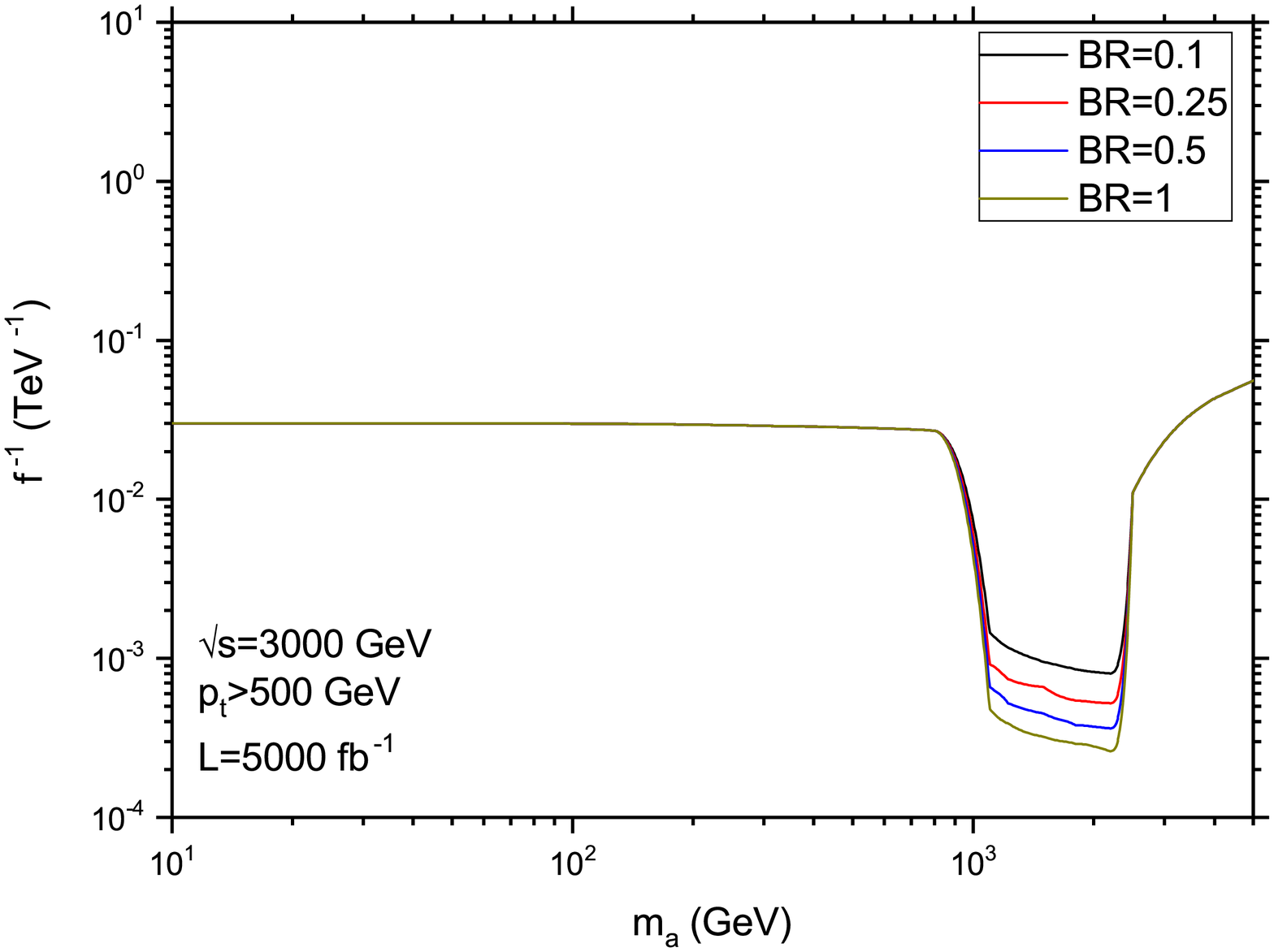}
\caption{The as in Fig.~\ref{fig:CBSS750}, but for $\sqrt{s} = 3000$
GeV and $L = 5000$ fb$^{-1}$.}
\label{fig:CBSS1500}
\end{center}
\end{figure}
%
\begin{figure}[htb]
\begin{center}
\includegraphics[scale=0.8]{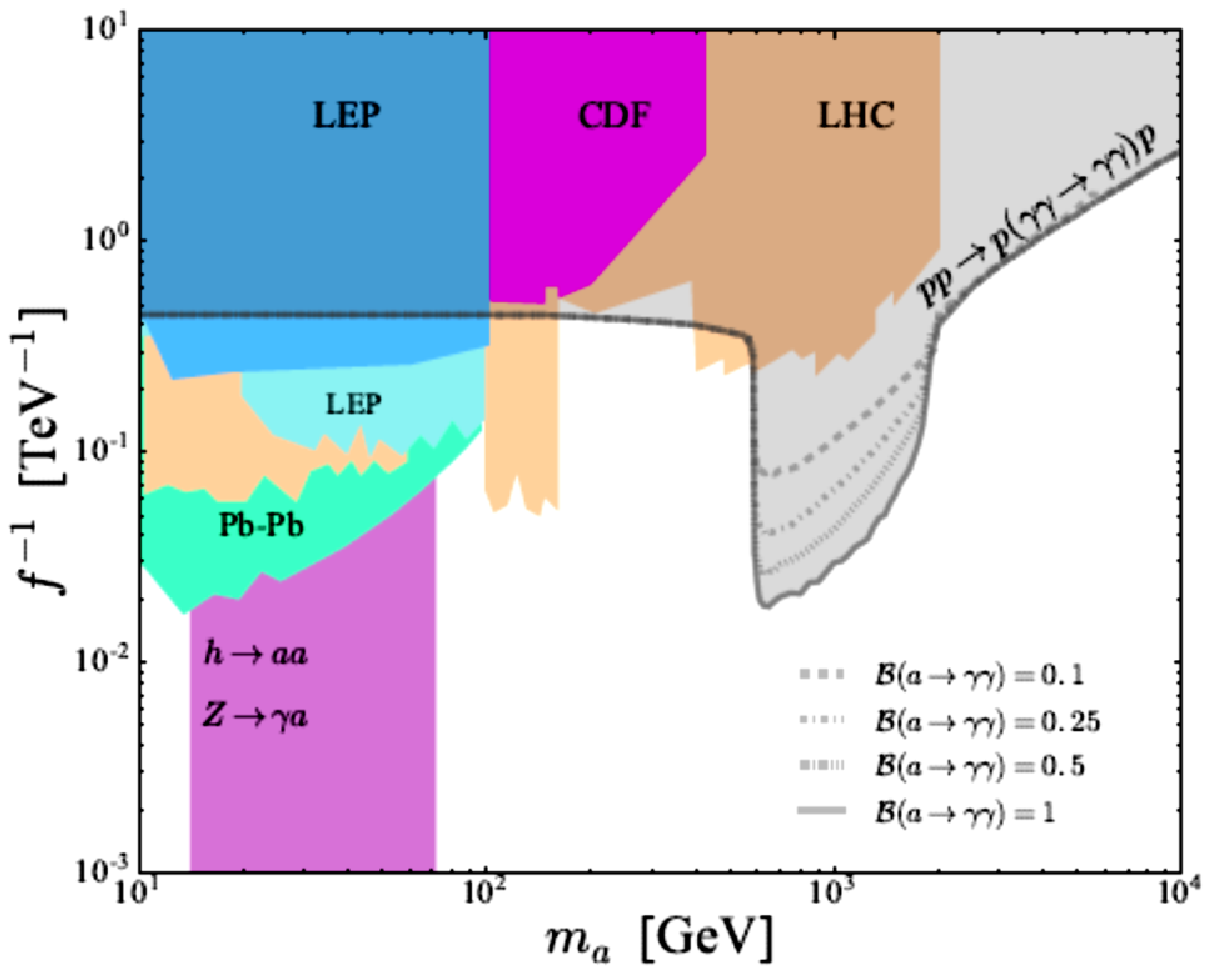}
\caption{The $95\%$ C.L. current exclusion regions for different
values of $\mathrm{Br}(a \rightarrow \gamma\gamma)$
\cite{Beldenegro:2018}.} \label{fig:balfig7}
\end{center}
\end{figure}

The sensitivity region in Fig.~\ref{fig:CBSS1500} is limited to a
rather sharp region. It is due to the fact that the cross section is
large only in the mass range $1000-2500$ GeV, see
Fig.~\ref{fig:CBMAE1500}. In this region the axion term dominates,
while outside it the contributions from the axion and SM terms are
comparable and they partially cancel each other. By comparison the
exclusion region in Fig.~\ref{fig:CBSS1500} with that for the
process $pp \rightarrow p(\gamma\gamma \rightarrow \gamma\gamma)p$
in Fig.~\ref{fig:balfig7}, we have to conclude that the above
mentioned behavior of the sensitivity is really related to a
specific dependence of the subprocess $\gamma\gamma \rightarrow
\gamma\gamma$ on the axion mass as discussed in Appendix~A.

\subsection{Weizs\"{a}cker-Williams photons} %

The photon-photon interactions can be realized at the CLIC in a
different way, using almost real photons emitted from incoming
electron beams. These processes can be studied in the framework of
the Weizs\"{a}cker-Williams approximation (WWA).
\cite{Weizsacker:1934}-\cite{Carimalo:1979}.

In this approximation, incoming electrons scatter at very small
angles. Therefore, electrons can not be caught in the main detector.
If scattered electrons of the beams are observed, minimal and
maximal values of scattered photon energies can be determined.
Otherwise, energy or momentum cuts imposed on final state particles
can be used to specify the minimum photon energy. The photon
virtuality varies in the following kinematical range
\begin{equation}\label{Qmin}
Q_{\min}^2 \leq Q^2 \leq Q_{\max}^2 \;,
\end{equation}
where $Q^2=-q^2$ ($q$ is the photon momentum), and
\begin{equation}\label{Qmm}
Q_{\min}^2=\frac{m_{e}^2 x^2}{1-x},  \;, \quad
Q_{\max}=\frac{4E_{e}^2}{1-x} \;.
\end{equation}
Here $E_e$ is the energy of the incoming electron, $E_\gamma$ is the
photon energy and $x=E_\gamma/E_e$. The scattered electrons have
very small scattering angles from the beam direction, and
their transverse momenta are small. Therefore, due to momentum
conservation, the transverse momenta of the emitted photons should
also be very small. This means that the virtuality of the photon
$Q^2$ in WWA is very small (almost real). In an experiment
performed by the DELPHI Collaboration, it was observed that the
virtualities of $90\%$ photons are less than $1$ GeV$^2$ using
appropriate experimental techniques \cite{delphi}. The WWA is also
useful for experimental studies due to it allows us to find cross
sections for the  process $e^{-}e^{+}\rightarrow e^{-}Xe^{+}$ via
subprocess $\gamma \gamma\rightarrow X $ \cite{Baur:2002}. In the
literature, there are many papers on photon-induced processes, see,
for instance, \cite{delphi}, \cite{phe1}-\cite{Ozguven:2017}.
%
\begin{figure}[htb]
\begin{center}
\includegraphics[scale=0.35]{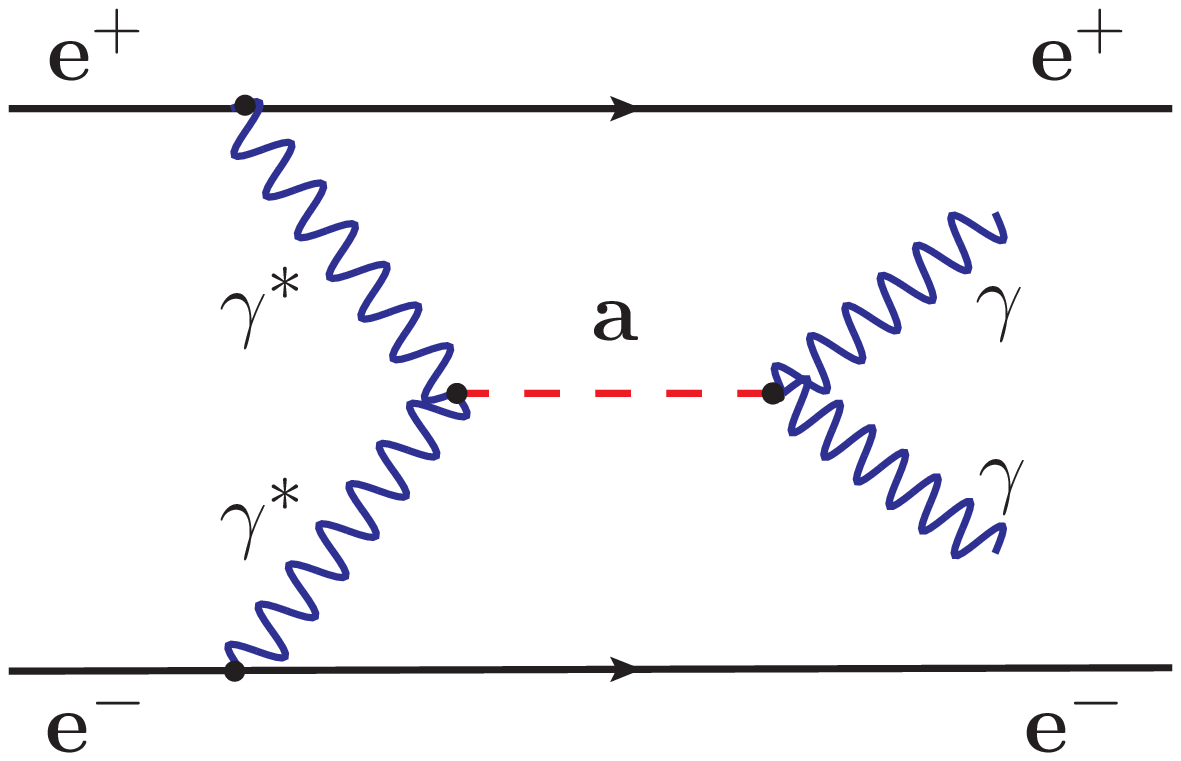}
\includegraphics[scale=0.35]{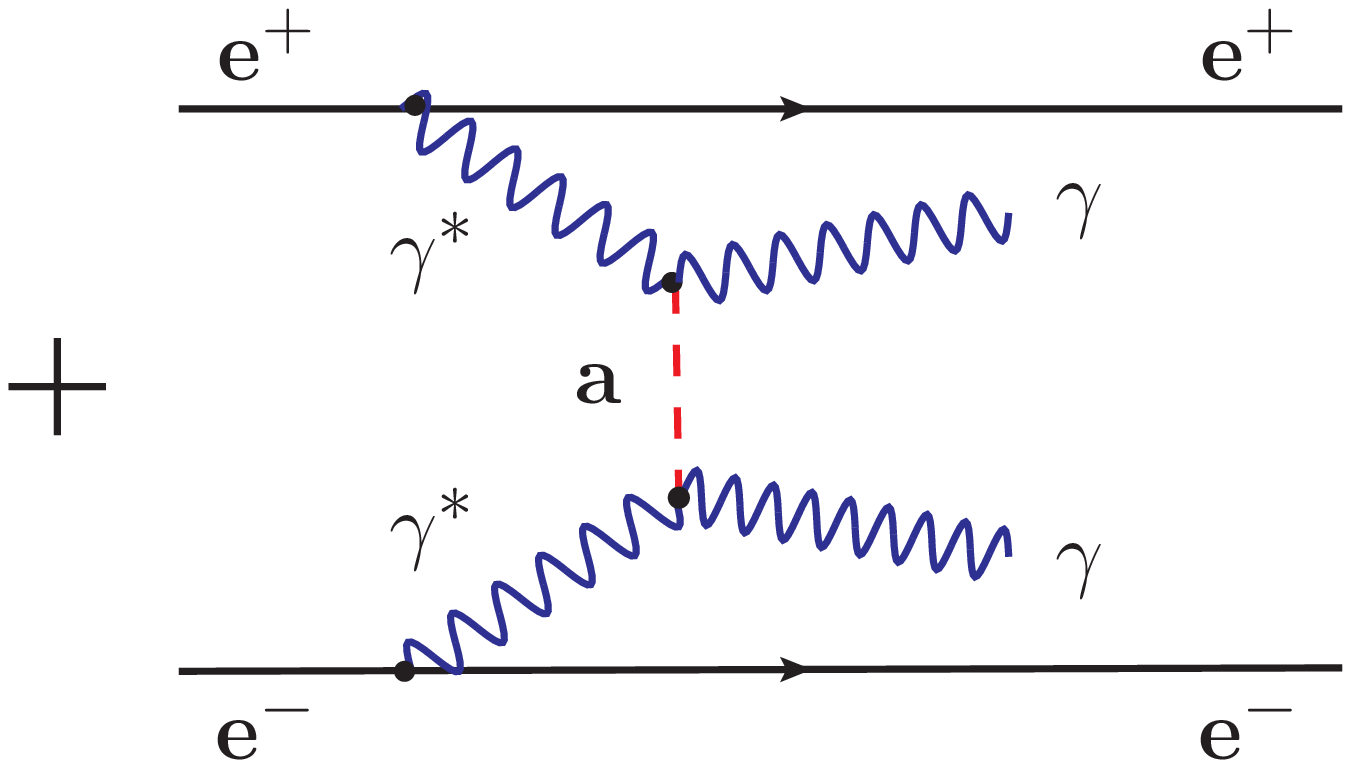}
\includegraphics[scale=0.35]{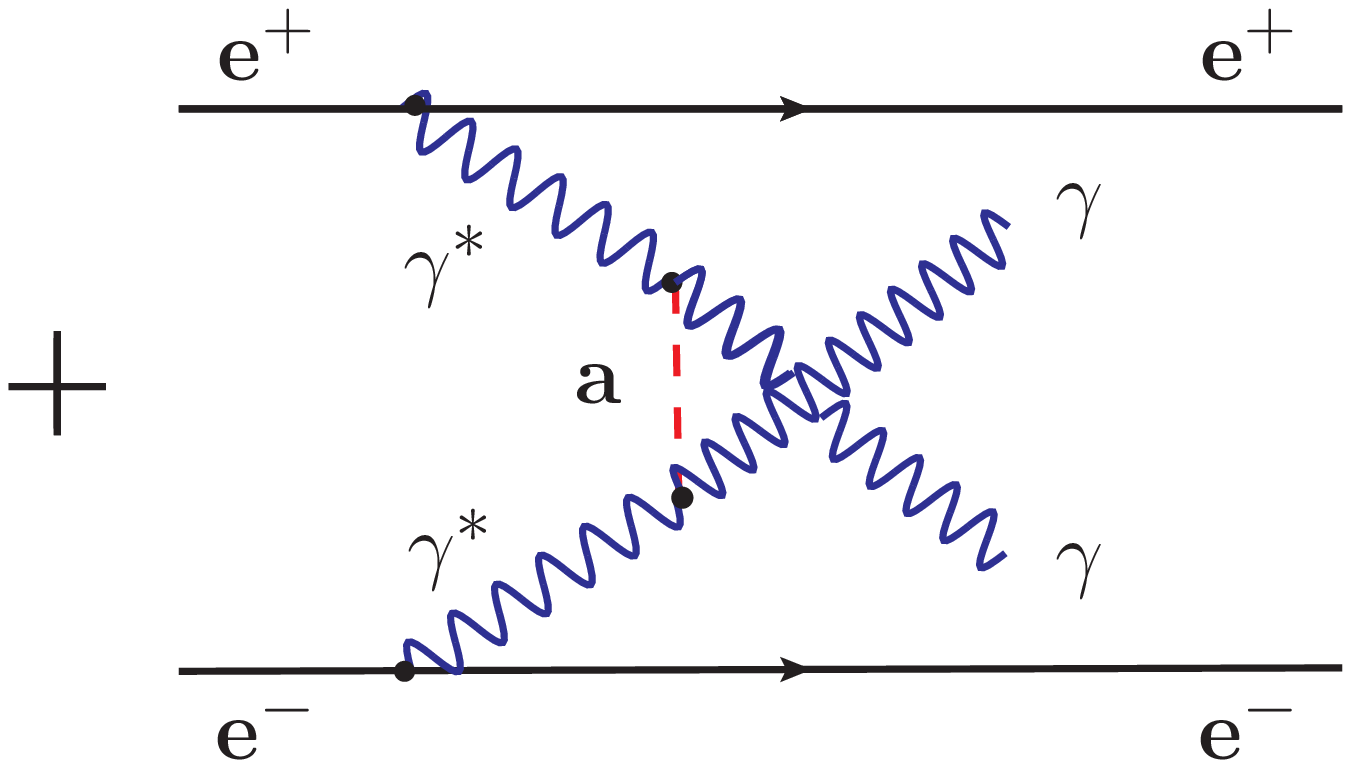}
\caption{The Feynman diagrams describing photon-induced light-by-light
virtual production of the axion-like particle $a$ in $e^+e^-$
collision.}
\label{fig:diagram}
\end{center}
\end{figure}

In the WWA, the photons have the following spectrum
\begin{eqnarray}\label{WW_spectrum}
f_{\gamma/e}(x) = \frac{\alpha}{\pi}\left[\left(\frac{1-x+x^2/2}{x}\right)\log\frac{Q_{\max}^2}
{Q_{\min}^2}-\frac{m_e^2x}{Q_{\min}^2}\left(1-\frac{Q_{\min}^2}{Q_{\max}^2}
\right) \right].
\end{eqnarray}
Here $m_{e}$ is the electron mass, $x=E_\gamma/E_e$ is the ratio of
the photon energy and energy of the incoming electron, and $\alpha$
is the fine structure constant. The cross section of the process
$e^{-}e^{+}\rightarrow e^{-}\gamma\gamma e^{+}$ can be calculated
using formula \eqref{cs} with the replacement $y_{\max} = z_{\max} =
1 - m_e/E_e$. In the light of above arguments,
in what follows, we take $Q_{\max}^2=2$ GeV$^2$.

In addition to the backgrounds mentioned in subsection 2.1, possible
backgrounds also came from $\gamma \gamma \rightarrow e^+e^- \gamma
\gamma$ and $ZZ$-induced processes. The first one was estimated in
\cite{ATLAS_ions_2} to be below 1\%. The second one may not be taken
into account since the $ZZ$ luminosity is approximately 100 times
smaller than the $\gamma\gamma$ luminosity \cite{ZZ_lum}.

\begin{figure}[htb]
\begin{center}
\includegraphics[scale=0.65]{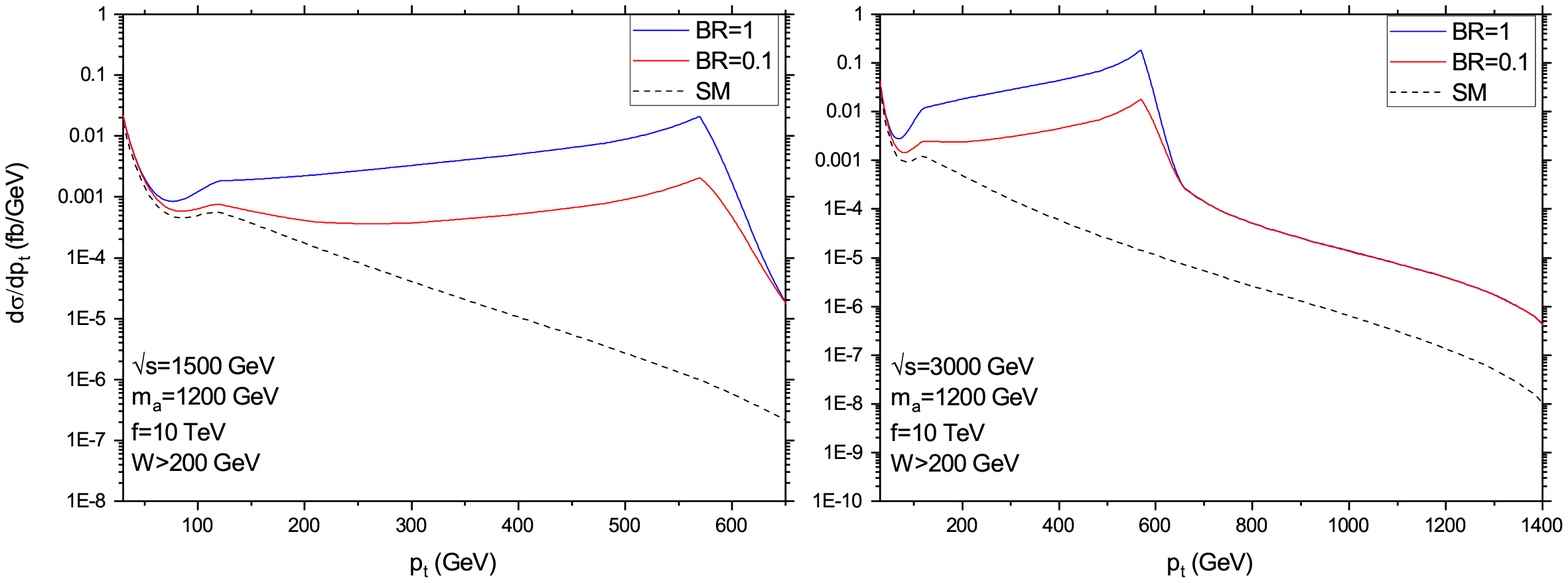}
\caption{The differential cross sections for the process $e^+e^-
\rightarrow e^+ \gamma\gamma e^- \rightarrow e^+ a \,e^- \rightarrow
e^+ \gamma\gamma e^-$ at the CLIC in the WWA for the initial photons
in the subprocess $\gamma\gamma \rightarrow \gamma\gamma$ as
functions of the transverse momenta of the final photons $p_t$ for
the ALP mass $m_a = 1200$ GeV and coupling constant $f = 10$ TeV.
The invariant energy is equal to $\sqrt{s} = 1500$ (3000) GeV on the
left (right) panel. The curves for both $\mathrm{Br}(a \rightarrow
\gamma\gamma) = 1.0$ and $\mathrm{Br}(a \rightarrow \gamma\gamma) =
0.1$ are shown. The dashed lines denote the SM contributions.}
\label{fig:WWPTD}
\end{center}
\end{figure}

\begin{figure}[htb]
\begin{center}
\includegraphics[scale=0.65]{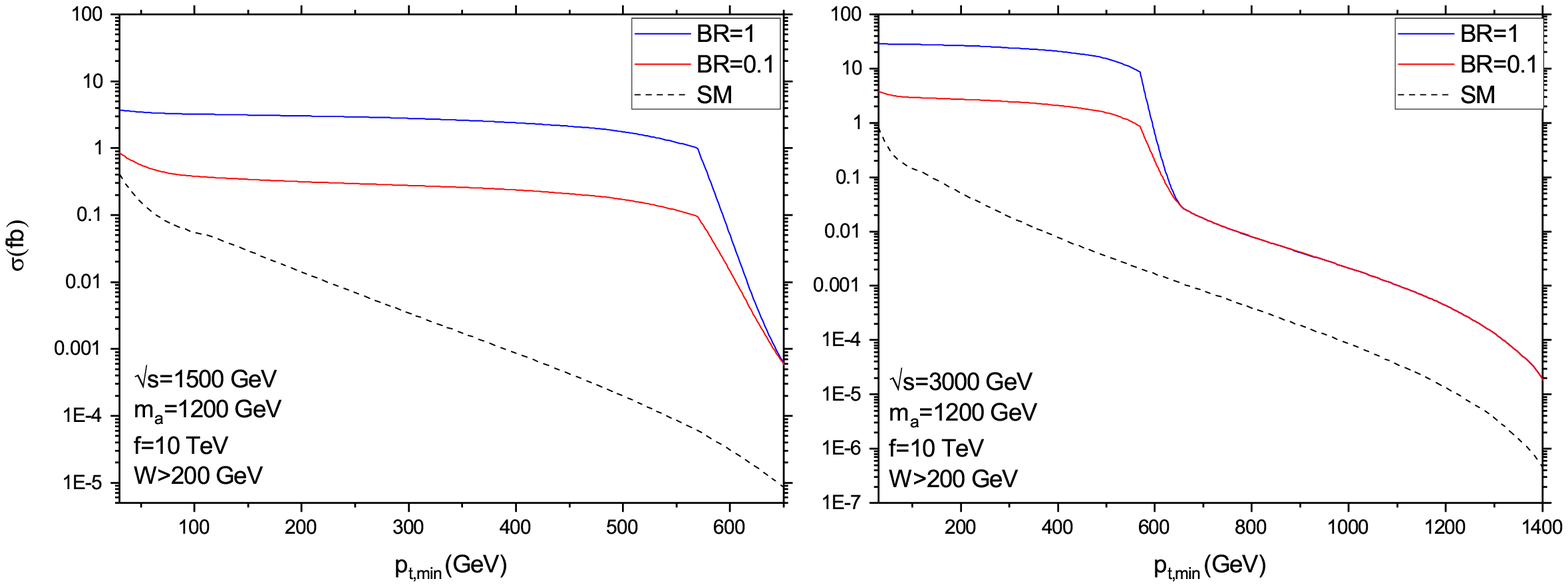}
\caption{The same as in Fig.~\ref{fig:WWPTD}, but for the total
cross sections as functions of the transverse momenta cutoff
$p_{\mathrm{t,min}}$ of the final photons.}
\label{fig:WWPTCUT}
\end{center}
\end{figure}

The results of our calculations of the differential and total cross
sections are presented in Figs.~\ref{fig:WWPTD} and
\ref{fig:WWPTCUT}. They should be compared with the cross sections
for the process induced by the CB photons shown in
Figs.~\ref{fig:CBPTD} and \ref{fig:CBPTCUTWCUT}. The WWA cross
sections have appeared to be approximately $10^4$ ($10^2$) times
smaller that the CB cross sections for $\sqrt{s} = 1500$ GeV
($\sqrt{s} = 3000$ GeV).
%
%
%

\begin{figure}[htb]
\begin{center}
\includegraphics[scale=0.65]{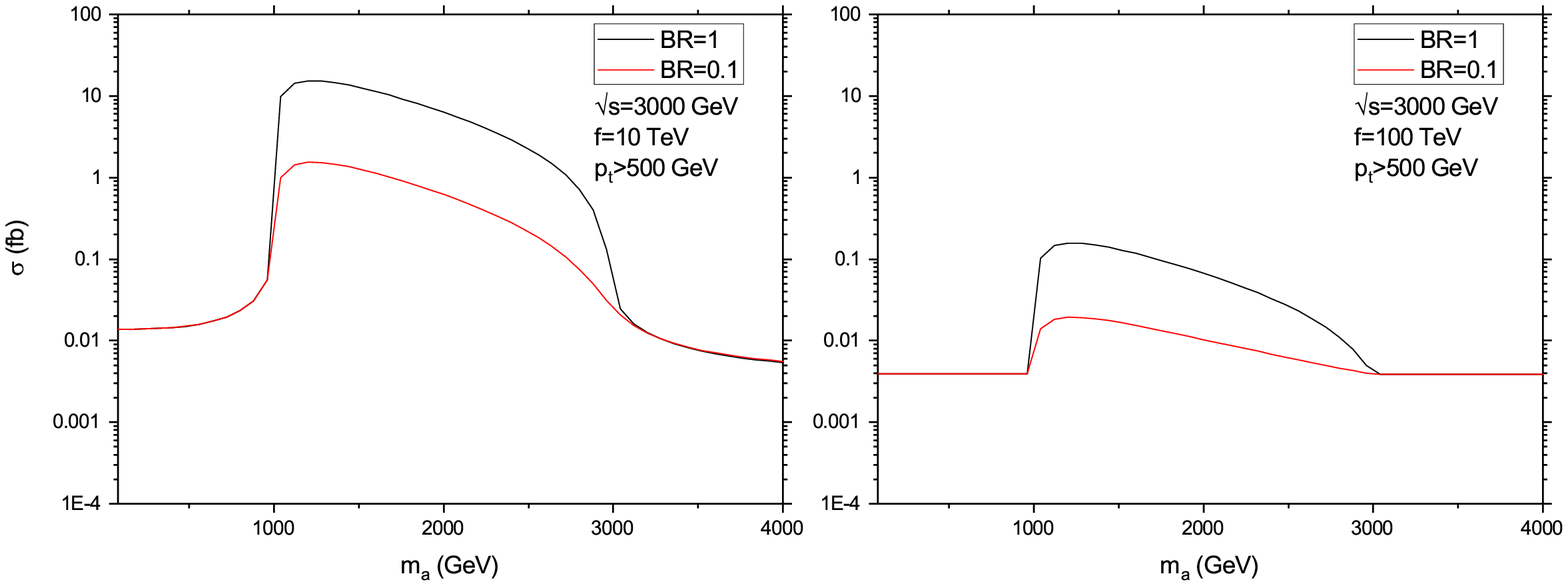}
\caption{The total cross sections for the process $e^+e^-
\rightarrow e^+ \gamma\gamma e^- \rightarrow e^+ a \,e^- \rightarrow
e^+ \gamma\gamma e^-$ at the CLIC in the WWA for the initial photons
as functions of the ALP mass $m_a$ for $f = 10$ TeV and $f = 100$
TeV with two values of $\mathrm{Br}(a \rightarrow \gamma\gamma)$.}
\label{fig:WWMAE1500}
\end{center}
\end{figure}

The same one can see in Fig.~\ref{fig:WWMAE1500}, where the total
cross section for the process $e^+e^- \rightarrow e^+ \gamma\gamma
e^- \rightarrow e^+ a \,e^- \rightarrow e^+ \gamma\gamma e^-$ is
shown as a function of the ALP mass $m_a$. For the ALP branching
ratio $\mathrm{Br}(a \rightarrow \gamma\gamma) = 1.0$, there are big
bumps in the curves in the mass region 1000 GeV--3000 GeV for both
values of the collision energy $\sqrt{s}$.
%

\begin{figure}[htb]
\begin{center}
\includegraphics[scale=0.55]{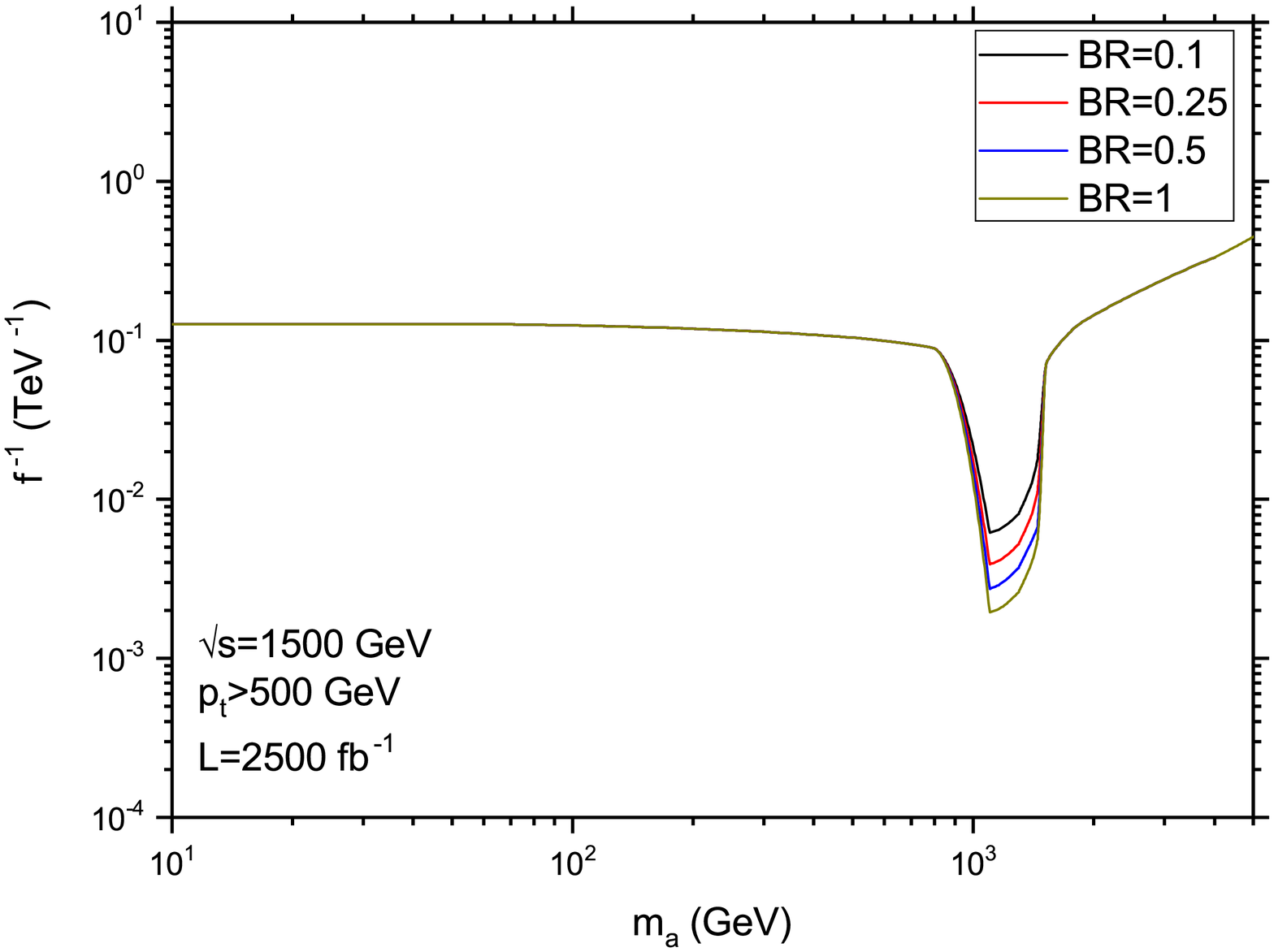}
\caption{The 95\% C.L. CLIC exclusion region for the process $e^+e^-
\rightarrow e^+ \gamma\gamma e^- \rightarrow e^+ a \,e^- \rightarrow
e^+ \gamma\gamma e^-$ with the invariant energy $\sqrt{s} = 1500$
GeV, transverse momentum cut on the final photons $p_t=500$ GeV,
integrated luminosity $L = 2500$ fb$^{-1}$, and different values of
$\mathrm{Br}(a \rightarrow \gamma\gamma)$. The WWA for the initial
photons in the subprocess $\gamma\gamma \rightarrow \gamma\gamma$ is
used.} \label{fig:WWSS750}
\end{center}
\end{figure}

\begin{figure}[htb]
\begin{center}
\includegraphics[scale=0.55]{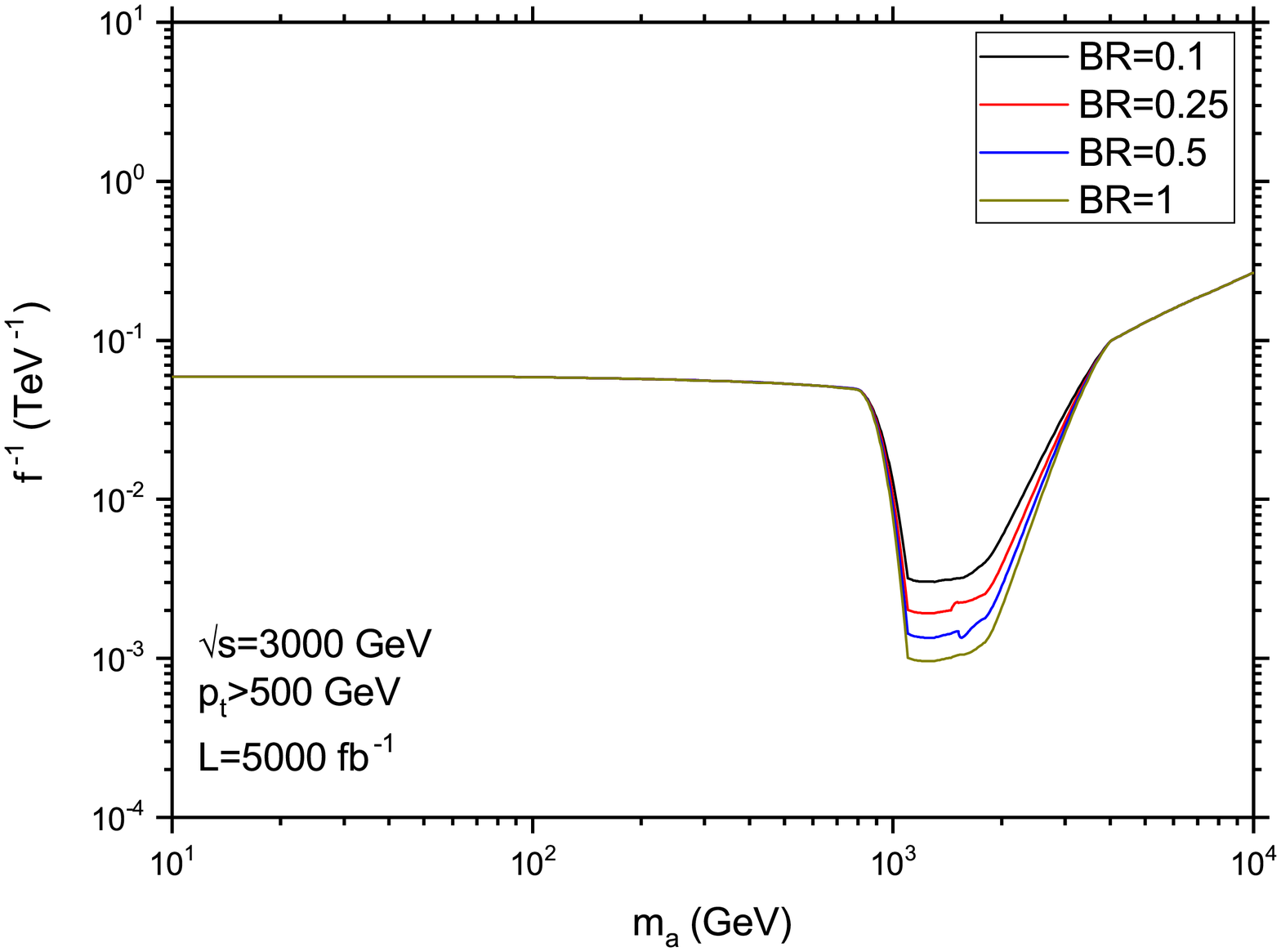}
\caption{The same as in Fig.~\ref{fig:WWSS750}, but for $\sqrt{s} =
3000$ GeV and $L = 5000$ fb$^{-1}$.}
\label{fig:WWSS1500}
\end{center}
\end{figure}

Fig.~\ref{fig:WWSS750} gives the $95\%$ C.L. CLIC exclusion region
in the $(m_a, f^{-1})$ plane in the case when the subprocess
$\gamma\gamma \rightarrow \gamma\gamma$ is induced by the WWA
photons with $\sqrt{s}=1500$ GeV and $L=2500$ fb$^{-1}$. As one can
see from this figure, the bounds are of the order of $10^{-1}$
TeV$^{-1}$ in the mass regions 10 GeV--1000 GeV. In the narrow mass
region 1000 GeV--1500 GeV it is obtained to be of the order of
$10^{-3}$ TeV$^{-1}$. Similarly, Fig.~\ref{fig:WWSS1500} shows the
$95\%$ C.L. exclusion region in the $(m_a, f^{-1})$ plane for
$\sqrt{s}=3000$ GeV and $L=5000$ fb$^{-1}$. In the mass region $10$
GeV-$1000$ GeV the bounds on $f^{-1}$ are of the order of $10^{-1}$
TeV$^{-1}$. In the mass range 1000 GeV--1500 GeV, these bounds reach
the value $1\times 10^{-3}$ TeV$^{-1}$. For both $\sqrt{s}$, the
bounds are much weaker than those for the CB initial photons.

%

\section{Conclusions} %

We have studied the possibility to search for heavy axion-like
particles in the process $\gamma \gamma \rightarrow \gamma \gamma $
with Compton backscattered initial photons and process  $e^+e^-
\rightarrow e^+e^- \gamma \gamma$ induced by light-by-light
scattering with Weizs\"{a}cker-Williams initial photons at the CLIC.
The calculations were made for the collision energy $\sqrt{s} =
1500$ GeV (2nd stage of the CLIC) and integrated luminosity $L =
2500$ fb$^{-1}$, as well as for the energy $\sqrt{s} = 3000$ GeV and
integrated luminosity $L = 5000$ fb$^{-1}$ (3rd stage of the CLIC).
It was assumed that the pseudoscalar ALP interacts with photons via
CP-even term in the Lagrangian \eqref{axion_photon_lagrangian}.

We 95\% C.L. exclusion regions in the plane ($m_a$, $f^{-1}$), where
$m_a$ is the ALP mass, $f^{-1}$ ALP--photon coupling, are given. The
results are presented for two values of $\sqrt{s}$ and $L$ as
functions of the ALP branching ratio into photons Br($a \rightarrow
\gamma\gamma$). The best bounds are obtained for Br$(a \rightarrow
\gamma\gamma) = 1$. Our calculations have shown that the numerical
results remain almost the same if we take into account the CP-odd
term instead of the CP-even one in the Lagrangian
\eqref{axion_photon_lagrangian}, with the same coupling $f^{-1}$.

By comparing our exclusion regions with other collider exclusion
regions, we may conclude that the ALP search at the CLIC has the
great physics potential of searching for the ALPs, especially, in
the mass region 1 TeV -- 2.4 TeV, for the collision energy $\sqrt{s}
= 3000$ GeV and integrated luminosity $L = 5000$ fb$^{-1}$. In
particular, our bounds are much stronger than recently obtained
bounds for the ALP virtual production in the process
$p(\gamma\gamma\rightarrow\gamma\gamma)p$ at the LHC
\cite{Beldenegro:2018}.



\setcounter{equation}{0}
\renewcommand{\theequation}{A.\arabic{equation}}

\section*{Appendix A}
\label{app:A}

Here we obtain an approximate formula for the cross section with the
CB initial photons. As it was already mentioned in the text, in the
mass region $1000-2500$ GeV the dominant contribution to the cross
section comes from the $s$-channel terms in $M_a$. Let us put $M =
M_{++++}$ \eqref{M++++}
\begin{equation}\label{M}
M = -\frac{4}{f^2} \frac{s^2}{s - m_a^2 +i m_a\Gamma_a} \;.
\end{equation}
Then
\begin{equation}\label{M2}
|M|^2 = \frac{16}{f^4} \frac{s^4}{(s - m_a^2)^2 + m_a^2\Gamma_a^2}
\;.
\end{equation}
We get from \eqref{M2} that at high energy
\begin{equation}\label{M2_asym}
|M|^2 \big|_{s \gg m_a^2} \sim \frac{s^2}{f^4} \;.
\end{equation}
However, simple dimensional arguments are not valid in the most
important resonance region $s \sim m_a^2$ in which
\begin{equation}\label{M2_reson}
|M|^2 \big|_{s \sim m_a^2} \sim \frac{m_a^6}{f^4 \Gamma_a^2} \;.
\end{equation}
Since $M$ depends only on $s$, we find from the relation
\begin{equation}\label{dif_sub_cs}
\frac{d\hat{\sigma}(s)}{d\Omega} = \frac{|M|^2}{64\pi^2 s}
\end{equation}
that the cross section of the subprocess
$\gamma\gamma\rightarrow\gamma\gamma$ is equal to
\begin{equation}\label{subproc_cs}
\hat{\sigma}(s) = \frac{1}{16\pi s} \,|M|^2 \;.
\end{equation}
The integrations in eq.~\eqref{cs} can be rewritten as follows
\begin{equation}\label{integration}
2z \,dz \frac{dy}{y} = dx_1 dx_2 \sim \frac{dx_1}{x_1}
\frac{ds}{4E_e^2} \;.
\end{equation}
Then the cross section of the process under consideration is given
by the integral
\begin{equation}\label{proc_cs}
\sigma = \frac{1}{4E_e^2}\int \!\!ds \,\frac{1}{16\pi s} \,|M|^2 \;.
\end{equation}

Let us estimate the contribution to $\sigma$ from the resonance
region
\begin{equation}\label{region}
m_a^2 - C m_a \Gamma_a \leqslant s \leqslant m_a^2 + C m_a \Gamma_a
\;,
\end{equation}
where $C$ is a constant of order $\mathrm{O}(1)$. After introducing
variable $x = (s - m_a^2)/(m_a\Gamma_a)$, we get
\begin{equation}\label{cross_section_2}
\sigma = \frac{1}{f^2} \,\frac{m_a^2}{E_e^2}
\,\mathrm{Br}(a\rightarrow\gamma\gamma)\int_{-C}^{C} dx
\,\frac{1}{x^2 + 1} \;.
\end{equation}
In order to numerically evaluate the cross section, we put $C=1$.
Then we obtain the estimate
\begin{equation}\label{cs_final}
\sigma \simeq 0.39 \times 10^6 \,\frac{\pi }{\frac{}{}2}\left(
\frac{\mathrm{TeV}}{f} \right)^{\!2} \left(
\frac{m_a}{E_e}\right)^{\!2} \mathrm{Br}(a\rightarrow\gamma\gamma)
\mathrm{\ fb} \;.
\end{equation}
Comparing this formula with Fig.~4, we see that it gives a correct
dependence of the cross section on the parameters $f$, $m_a$ and
$\mathrm{Br}(a\rightarrow\gamma\gamma$ in the mass region
$1100-2500$ GeV. As for numerical values of $\sigma$, we find, for
example, for $E_e = 3.0$ TeV, $f = 10$ TeV, $m_a = 1.5$ TeV, and
$\mathrm{Br}(a\rightarrow\gamma\gamma) = 1.0$
\begin{equation}\label{cs_num}
\sigma \simeq 1.6 \times 10^3 \mathrm{\ fb} \;.
\end{equation}
An accurate integration with an account of the photon spectra should
modify the numerical factor in the right-hand-side of
eq.~\eqref{cs_final}. This formula must be regarded only as an
illustration of how the main contribution to the cross section comes
from the resonance region of the pure axion amplitude.




\end{document}